\documentstyle[12pt]{article}

\newcommand{\be}{\begin{equation}}
\newcommand{\ee}{\end{equation}}

\newcommand{\ra}{\rightarrow}

\newcommand{\ep}{\varepsilon}

\newcommand{\lbd}{\lambda}

\begin{document}

\begin{center}
{\large{\bf Bose-Einstein condensation}} \\ [5mm]
{\large{\bf V.I. Yukalov$^{1,2}$}} \\ [3mm]
{\it $^1$Bogolubov Laboratory of Theoretical Physics, \\
Joint Institute for Nuclear Research, Dubna 141980, Russia\\ [2mm]
$^2$Institut f\"ur Theoretische Physik, \\
Freie Universit\"at Berlin, Arnimallee 14, D-14195 Berlin, Germany}
\end{center}

\vskip 10mm

Bose-Einstein condensation is the {\it occupation of a single quantum
state by a large number of identical particles}. This implies that particles
are assumed to be bosons, satisfying the Bose-Einstein statistics allowing
for many particles to pile up in the same quantum state. This is contrary
to fermions, satisfying the Fermi-Dirac statistics, for which the Pauli
exclusion principle does not allow the occupation of any single quantum
state by more than one particle.

The role of quantum correlations, caused by the Bose-Einstein statistics, 
is crucial for the occurrence of BEC. This statistics was advanced by 
Bose (1924) for photons, having zero mass, and generalized by Einstein 
(1924) to particles with nonzero masses. Einstein (1925) also described 
the phenomenon of condensation in ideal gases. The possibility of BEC in 
weakly nonideal gases was theoretically demonstrated by Bogolubov (1947). 
The wave function of Bose-condensed particles in dilute gases satisfies 
the Gross-Pitaevskii equation, suggested by Gross (1961) and Pitaevskii 
(1961). Its mathematical structure is that of the nonlinear Schr\"odinger 
equation. Experimental evidence of BEC in weakly interacting confined 
gases was achieved, seventy years after the Einstein prediction, almost 
simultaneously by three experimental groups (Anderson {\it et al.}, 1995; 
Bradley {\it et al.}, 1995; Davis {\it et al.}, 1995).

To say that many particles are in the same quantum state is equivalent to 
saying that these particles display the {\it state coherence}. That is, 
BEC is a particular case of {\it coherence phenomena} related to the 
arising state coherence. For the latter to occur, the particles are to be 
strongly correlated with each other. The expected conditions, when such a 
strong correlation could appear, may be qualitatively understood applying 
the de Broglie duality arguments to an ensemble of atoms in thermodynamic 
equilibrium at temperature $T$. Then the thermal energy of an atom is 
given by $k_BT$, where $k_B$ is the Boltzmann constant. This energy 
defines the thermal wavelength 
\be
\label{1} 
\lbd_T \equiv \sqrt{\frac{2\pi\hbar^2}{m_0k_BT}} 
\ee 
for an atom of mass $m_0$, with $\hbar$ being the Planck constant. Thus, an 
atom can be associated with a matter wave characterized by the wavelength 
(1). Atoms become correlated with each other when their related waves 
overlap, which requires that the wavelength (1) be larger than the mean 
interatomic distance, $\lbd_T>a$. The average atomic density $\rho\equiv N/V$ 
for $N$ atoms in volume $V$ is related to the mean distance $a$ through 
the equality $\rho a^3=1$. Hence, condition $\lbd_T>a$ may be rewritten as 
$\rho\lbd_T^3>1$. With the thermal wavelength (1), this yields the inequality
\be
\label{2}
T < \frac{2\pi\hbar^2}{m_0 k_B}\; \rho^{2/3} \; ,
\ee
which tells that state coherence may develop if temperature is sufficiently
low or the density of particles is sufficiently high.

An accurate description of BEC for an ideal gas is based on the {\it
Bose-Einstein distribution}
\be
\label{3} 
n({\bf p}) =\left [ \exp\left ( \frac{\ep_p-\mu}{k_B
T}\right ) - 1 \right ]^{-1} \; , 
\ee 
describing the density of particles with a single-particle energy 
$\ep_p=p^2/2m_0$ for a momentum ${\bf p}$ and with a chemical potential $\mu$.
The latter is defined from the condition $N=\sum_p n({\bf p})$ for the total
number of particles. Assuming the {\it thermodynamic limit} 
$$
N\ra \infty \; , \qquad V \ra\infty \; , \qquad \frac{N}{V} \ra const 
$$ 
allows the replacement of summation over ${\bf p}$ by integration. Then 
the fraction of particles, condensing to the state with ${\bf p}=0$, is
\be
\label{4}
n_0 \equiv \frac{N_0}{N} = 1 - \left ( \frac{T}{T_c} \right )^{3/2}
\ee
below the {\it condensation temperature}
\be
\label{5}
T_c = \frac{2\pi\hbar^2\rho^{2/3}}{m_0 k_B\zeta^{2/3}} \; ,
\ee
where $\zeta\approx 2.612$, while $n_0=0$ above the critical temperature (5).
The latter is about twice smaller than the right-hand side of inequality (2).

The condensate fraction (4) is derived for an ideal, that is, noninteracting, 
Bose gas. A weakly nonideal, i.e., weakly interacting Bose gas, also 
displays Bose-Einstein condensation, though particle interactions deplete 
the condensate, so that at zero temperature the condensate fraction is 
smaller than unity, $n_0<1$. A system is called weakly interacting if the 
characteristic interaction radius $r_{int}$ is much shorter than the mean 
interparticle distance, $r_{int}\ll a$. This inequality can be rewritten 
as $\rho r_{int}^3\ll 1$, because of which such a system is termed dilute.

Superfluid liquids, such as liquid $^4$He, are far from being dilute.
Nevertheless, one commonly believes that the phenomenon of superfluidity 
is somehow connected with BEC, although an explicit relation between the 
superfluid and condensate fractions is not known. Theoretical calculations 
and experimental observations for superfluid helium estimate the condensate 
fraction at $T=0$ as $n_0\approx 0.1$.

A strongly correlated pair of fermions can approximately be treated as a boson.
This is why the arising superfluidity in liquid $^3$He can be interpreted as
the condensation of coupled fermions. Similarly, superconductivity is often 
compared with the condensation of the Cooper pairs that are formed by 
correlated electrons or holes. However, one should understand that 
superconductivity of fermions is analogous but not identical to BEC of 
bosons.

An ideal object for the experimental observation of BEC is a dilute atomic
Bose gas confined in a trap and cooled down to temperatures satisfying
condition (2).  Such experiments with different atomic gases have been
recently realized, BEC explicitly observed, and a variety of its features
carefully investigated. It has been demonstrated that the system of 
Bose-condensed atoms displays a high level of state coherence.

There exist different types of traps, magnetic, optical, and their 
combinations, which makes it possible to confine atoms for sufficiently long 
times of up to 100 s. There are single-well and double-well traps. Employing 
a standing wave of laser light, one forms a multi-well periodic effective 
potentials called {\it optical lattices}. The use of the latter has allowed 
the demonstration of a number of interesting effects, as Bloch oscillations, 
Landau-Zener tunneling, Josephson current, Wannier-Stark ladders, Bragg 
diffraction, and so on.

An ensemble of Bose-condensed atoms, displaying a high level of state 
coherence, forms a matter wave that is analogous to a coherent electromagnetic
wave from a laser. Therefore a device emitting a coherent beam of Bose atoms 
is named {\it atom laser}.

The realization of BEC of dilute trapped gases is of great importance because 
of several reasons. First of all, this explicitly proved the actual existence 
of the phenomenon predicted by Einstein many years ago. It is worth 
emphasizing that a direct observation of BEC in superfluid helium, despite 
enormous experimental efforts, have never been achieved. This is why such an 
unambiguous observation of BEC of trapped atoms is of great value. Second, 
dilute atomic gases are sufficiently simple statistical systems, which 
can serve as a touchstone for testing different theoretical approaches. 
Finally, Bose-condensed trapped gases display such an enormous variability of 
their properties that they will certainly find diverse practical applications.

\vskip 1cm

{\it See also} Coherence phenomena; Critical phenomena; Lasers; Nonequilibrium 
statistical mechanics; Nonlinear optics; Nonlinear Schr\"odinger equations; 
Order parameters; Pattern formation; Phase transitions; Quantum nonlinearity; 
Quantum theory; Superconductivity; Superfluidity

\vskip 5mm

{\bf Further Reading}

\vskip 2mm

Anderson, M.H., Ensher, J.R., Matthews, M.R., Wieman, C.E. and Cornell, 
E.A. 1995. Observation of Bose-Einstein condensation in a dilute atomic 
vapor. {\it Science}, 269: 198--201 

\vskip 2mm

Bogolubov, N.N. 1947. On the theory of superfluidity. {\it Journal
of Physics} (Moscow), 11: 23--32

\vskip 2mm

Bose, S.N. 1924. Plancks gesetz und lichtquantenhypothese. {\it
Zeitschrift f\"ur Physik}, 26: 178--181

\vskip 2mm

Bradley, C.C., Sackett, C.A., Tollett, J.J. and Hulet, R.G. 1995.
Evidence of Bose-Einstein condensation in an atomic gas with
attractive interactions. {\it Physical Review Letters}, 75:
1687--1690

\vskip 2mm

Coleman, A.J. and Yukalov, V.I. 2000. {\it Reduced Density
Matrices}, Berlin: Springer

\vskip 2mm

Courteille, P.W., Bagnato, V.S. and Yukalov, V.I. 2001.
Bose-Einstein condensation of trapped atomic gases. {\it Laser
Physics}, 11: 659--800

\vskip 2mm

Dalfovo, F., Giorgini, S., Pitaevskii, L.P. and Stringari, S.
1999. Theory of Bose-Einstein condensation in trapped gases. {\it
Reviews of Modern Physics}, 71: 463--512

\vskip 2mm

Davis, K.B., Mewes, M.O., Andrews, M.R., van Drutten, N.J.,
Durfee, D.S., Kurn, D.M. and Ketterle, W. 1995. Bose-Einstein
condensation in a gas of sodium atoms. {\it Physical Review
Letters}, 75: 3969--3973

\vskip 2mm

Einstein, A. 1924. Quantentheorie des einatomigen idealen gases.
{\it Sitzungsberichte der Preussischen Akademie der Wissenschaften, Physik-
Mathematik}, 1924: 261--267

\vskip 2mm

Einstein, A. 1925. Quantentheorie des einatomigen idealen gases.
Zweite abhandlung. {\it Sitzungsberichte der Preussischen Akademie der 
Wissenschaften, Physik-Mathematik}, 1925: 3--14

\vskip 2mm

Gross, E.P. 1961. Structure of a quantized vortex in boson
systems. {\it Nuovo Cimento}, 20: 454--477.

\vskip 2mm

Huang, K. 1963. {\it Statistical Mechanics}, New York: Wiley

\vskip 2mm

Klauder, J.R. and Skagerstam, B.S. 1985. {\it Coherent States},
Singapore: World Scientific

\vskip 2mm

Lifshitz, E.M. and Pitaevskii, L.P. 1980. {\it Statistical
Physics: Theory of Condensed State}, Oxford: Pergamon

\vskip 2mm

Nozi\`eres, P. and Pines, D. 1990. {\it Theory of Quantum Liquids:
Superfluid Bose Liquids}, Redwood: Addison-Wesley

\vskip 2mm

Parkins, A.S. and Walls, D.F. 1998. The physics of trapped
dilute-gas Bose-Einstein condensates. {\it Physics Reports}, 303:
1--80

\vskip 2mm

Pitaevskii, L.P. 1961. Vortex lines in an imperfect Bose gas. {\it
Journal of Experimental and Theoretical Physics}, 13: 451--455

\vskip 2mm

Ter Haar, D. 1977. {\it Lectures on Selected Topics in Statistical
Mechanics}, Oxford: Pergamon

\vskip 2mm

Yukalov, V.I. and Shumovsky, A.S. 1990. {\it Lectures on Phase
Transitions}, Singapore: World Scientific

\vskip 2mm

Ziff, R.M., Uhlenbeck, G.E. and Kac, M. 1977. The ideal
Bose-Einstein gas revisited. {\it Physics Reports}, 32: 169--248

\end{document}